%% file: kdd_2014_prediction.tex
\documentclass{sig-alternate-2013}

\usepackage[utf8]{inputenc}
\usepackage{amsmath}
\usepackage{amsfonts}
\usepackage{mathtools}
\usepackage{xcolor}
\usepackage{booktabs}

\makeatletter
\newif\if@restonecol
\makeatother

\usepackage[ruled,lined,linesnumbered]{algorithm2e}

\def \calG {\mathcal{G}}
\def \calN {\mathcal{N}}
\def \calE {\mathcal{E}}
\def \calL {\mathcal{L}}

\newcommand{\mtrx}[1]{{#1}}
\newcommand{\vect}[1]{{#1}}

\newfont{\mycrnotice}{ptmr8t at 7pt}
\newfont{\myconfname}{ptmri8t at 7pt}
\let\crnotice\mycrnotice%
\let\confname\myconfname%

\permission{Permission to make digital or hard copies of all or part of this work for personal or classroom use is granted without fee provided that copies are not made or distributed for profit or commercial advantage and that copies bear this notice and the full citation on the first page. Copyrights for components of this work owned by others than the author(s) must be honored. Abstracting with credit is permitted. To copy otherwise, or republish, to post on servers or to redistribute to lists, requires prior specific permission and/or a fee. Request permissions from permissions@acm.org.
}
\conferenceinfo{SNAKDD'14, }{August 24-27 2014, New York City, NY, USA}
\copyrightetc{Copyright is held by the owner/author(s). Publication rights licensed to ACM.\\
\the\acmcopyr}
\crdata{978-1-4503-3192-0/14/08\ ...\$15.00.\\
http://dx.doi.org/10.1145/2659480.2659492}

\begin{document}

\title{Harnessing Mobile Phone Social Network Topology to Infer Users Demographic Attributes}

\numberofauthors{4}

\author{
\alignauthor
Jorge Brea\\
       \affaddr{Grandata Labs}\\
       \affaddr{Buenos Aires, Argentina}\\
       \texttt{jorge@grandata.com}
\alignauthor
Javier Burroni\\
       \affaddr{Grandata Labs}\\
       \affaddr{Buenos Aires, Argentina}\\
       \texttt{javier.burroni@grandata.com}
\alignauthor
Martin Minnoni\\
       \affaddr{Grandata Labs}\\
       \affaddr{Buenos Aires, Argentina}\\
       \texttt{martin@grandata.com}
\and
\alignauthor
Carlos Sarraute\titlenote{Corresponding author.}\\
       \affaddr{Grandata Labs}\\
       \affaddr{Buenos Aires, Argentina}\\
       \texttt{charles@grandata.com}
}

\date{}

\maketitle

\begin{abstract}

We study the structure of the social graph of mobile phone users in the country
of Mexico, with a focus on demographic attributes of the users
(more specifically the users' age).
We examine assortativity patterns in the graph, and observe
a strong age homophily in the communications preferences.
We propose a graph based algorithm for the prediction of the
age of mobile phone users.
The algorithm exploits the topology of the mobile phone network, together with a subset
of known users ages (seeds), to infer the age of remaining users. 
We provide the details of the methodology, and show experimental results
on a network $\calG_T$ with more than $70$ million users.
By carefully examining the topological
relations of the seeds to the rest of the nodes in $\calG_T$, 
we find topological metrics which have a direct influence on the performance
of the algorithm.
In particular we characterize subsets of users for which the accuracy of the algorithm 
is $62\%$ when predicting between 4 age categories
(whereas a pure random guess would yield an accuracy of $25\%$). 
We also show that we
can use the probabilistic information computed by the algorithm
to further increase its
inference power to $72\%$ on a significant subset of users.

\end{abstract}


\input{introduction}

\input{methods}

\input{results}

\input{conclusion}


\bibliographystyle{abbrv}
\bibliography{sna}

\end{document}

%% file: introduction.tex
\section{Introduction}

Over the past decade, mobile phones have become prevalent in all
parts of the world, across all demographic backgrounds. Mobile phones
are used by men and women across a wide age range in both
developed and developing countries.
They have become one of the most important mechanisms for
social interaction within a population, making them an increasingly
important source of information to understand human behavior,
human demographics, and the correlations between them. 
In particular, mobile phone usage information is being used to perform quantitative analysis
on the demographics of users regarding their age group, gender, education, and socioeconomic 
status \cite{blumenstock2010mobile,blumenstock2010s,frias2010gender}.

In this work we combine two sources of information: 
communication logs from a major mobile operator in Mexico,
and information on the demographics of a subset of the users
population.
This allowed us to first perform an observational study of 
mobile phone usage, differentiated by categories such as gender and age groups \cite{sarraute2014}.
This study is interesting in its own right, 
since it provides knowledge on the structure and
demographics of the mobile phone market in Mexico.
Here we focus our study on one variable: the age of the users.

Furthermore, we tackle the problem of inferring demographic attributes, such as their age.
The ability to predict such features has numerous applications for 
social organizations (for example, enabling them to perform specific health campaigns for women \cite{frias2010gender})
and for businesses, allowing them to gain insights on their customers and on the open market,
to segment (current and potential) customers by demographic attributes, 
and to perform directed marketing campaigns.

A semantic description of the network may consist
in characterizing particular user attributes (e.g. gender, age or credit risk),
or link attributes describing relationships between users (e.g. friendship, family or
working relationship).
One can also characterize individual users based on
more abstract social attributes derived from their communication patterns~\cite{frias2010gender}
or their geographic mobility \cite{gonzalez2008understanding,ponieman2013human}, and study how these features relate to other social and behavioral properties \cite{cho2011friendship}.
Methods such as logistic regression, readily used in this approach, emphasize finding similarity patterns
between nodes given their known features and mostly rely on the assumption that the selected node features
have a functional relationship with the particular attribute one wants to infer.

A complementary approach is to focus on the complex
relationships between individuals given by their social network, and exploit both short and long range interactions to infer feature relationships
among all individuals in the network. Methods such as Laplacian spectral decomposition and graph
diffusion algorithms, which emphasize the topological relationship between nodes within the network, 
have been widely used~\cite{Chapelle2006}. These methods rely on the existence of social or group properties such as homophily and other community structures prevalent in the network for a successful implementation.

Our main contribution in this work is to present an exclusively graph based inference method, 
relying only on the
topological structure of the mobile users network, together with a topological analysis of the 
performance of the algorithm.
The equations of our algorithm can be described as a diffusion process with two added properties: 
(i) memory of its initial state, and (ii) the information is propagated as a probability vector 
for each node attribute (instead of the value of the attribute itself).  
We show that these two additions to the diffusion process significantly improve its inference power in the
mobile network studied.
Our algorithm can successfully infer different
age groups within the network population given known values for a subset of nodes
(the seed set). Most interestingly, we show
that by carefully analyzing the topological relationships between correctly 
predicted nodes and the seed nodes, we can characterize particular subsets
of nodes for which our inference method has significantly higher accuracy, going up to 62\% (when predicting between 4 age categories). 
We then show that we can use the information in the probability vector of each node to
again increase our predictive power (up to 72\%).

The remainder of this paper is organized as follows.
In section \ref{sec:datasource} we provide an overview
of the datasets used in this study.
We also describe the observations performed on the social graph, in particular concerning  
age homophily among users.
In section~\ref{sec:methods} we describe our graph based algorithm and the topological metrics 
used in the analysis.
In section~\ref{sec:results} we present the results of the inference power of our algorithm.
We focus on understanding how the performance of the algorithm depends on age groups and the topological
properties of correctly inferred nodes. We then reexamine our results on the subnetwork composed
solely of the mobile operator's clients.
We further study how the model parameters and the added information contained in the probability vectors
can enhance the inference power of the algorithm.
Finally in section \ref{sec:conclusion} we conclude this
work remarking the most significant findings and ideas for future work.

\section{Data Source and Observations} \label{sec:datasource}

\subsection{Dataset Description} \label{sec:dataset}

The dataset used in this work consists of cell phone calls and SMS (short message
service) records (CDRs) for a three month period collected
by a mobile phone operator for the whole country of Mexico. 
Each record contains
detailed information about the anonymized phone numbers of both the caller and
the callee, whether it is an incoming or outgoing call, the duration and time stamp of
the call, and the location of the antenna servicing the client involved in the
communication. 
The detailed information provided by this dataset allows for the
construction of large scale complex dynamic networks with added node (client)
mobility based on the antenna location, which can be used for the study of
human mobility \cite{ponieman2013human}.

A milestone of this work is to study how
the ``bare bones'' structure of the mobile phone network contains useful information for the
 inference of the users age group. For our purposes, we summarize each
call/SMS record as a tuple $(i,j)$ representing a call/SMS between users $i$ and $j$.
We then aggregate these tuples for all calls/SMS for a three
month period into an edge list $(n_i,n_j,w_{i,j})$ where nodes $n_i$ and $n_j$ 
represent users $i$ and $j$ respectively and $w_{i,j}$ is a boolean value
indicating whether these two users have communicated at least once within the 
three month period. This edge list will represent our mobile graph  
$\calG = \left< \calN, \calE \right> $ where $\calN$ denotes the set of nodes (users) 
and $\calE$ the set of communication links. We note that only a subset $\calN_C$ nodes in $\calN$
are clients of the mobile operator, the remaining nodes $\calN \setminus \calN_C$ are
users that communicated with users in $ \calN_C $ but themselves are not clients of
the mobile operator. This distinction implies that our network $ \calG $ is an
incomplete network in the sense that communications among members in
$\calN \setminus \calN_C$ are not observed in $\calG$.

Together with the CDR records, we had access to several
demographic variables for over 500,000 clients. These demographic variables include
their actual age. We therefore distinguish three sets of nodes in our
graph $\calG$: the whole set $\calN$, the set of client nodes $\calN_C$, and the set
of client nodes with known age $\calN_{GT}$ which we use as the ground truth nodes
($\calN_{GT} \subset \calN_C \subset \calN$).
Lastly we split our set $\calN_{GT}$ into a set of seed nodes $\calN_S$ consisting of
approximately $75\%$ of $\calN_{GT}$ which we use as seeds to generate age predictions
for the remaining nodes in $\calG$, and the remaining nodes $\calN_V$ which we use as
a validation set to evaluate the performance of our predictions.

\begin{figure}[ht]
	\centering
	{\includegraphics[trim=1.2cm 0.2cm 2.0cm 1.2cm, clip=true, width=0.99\linewidth]
	{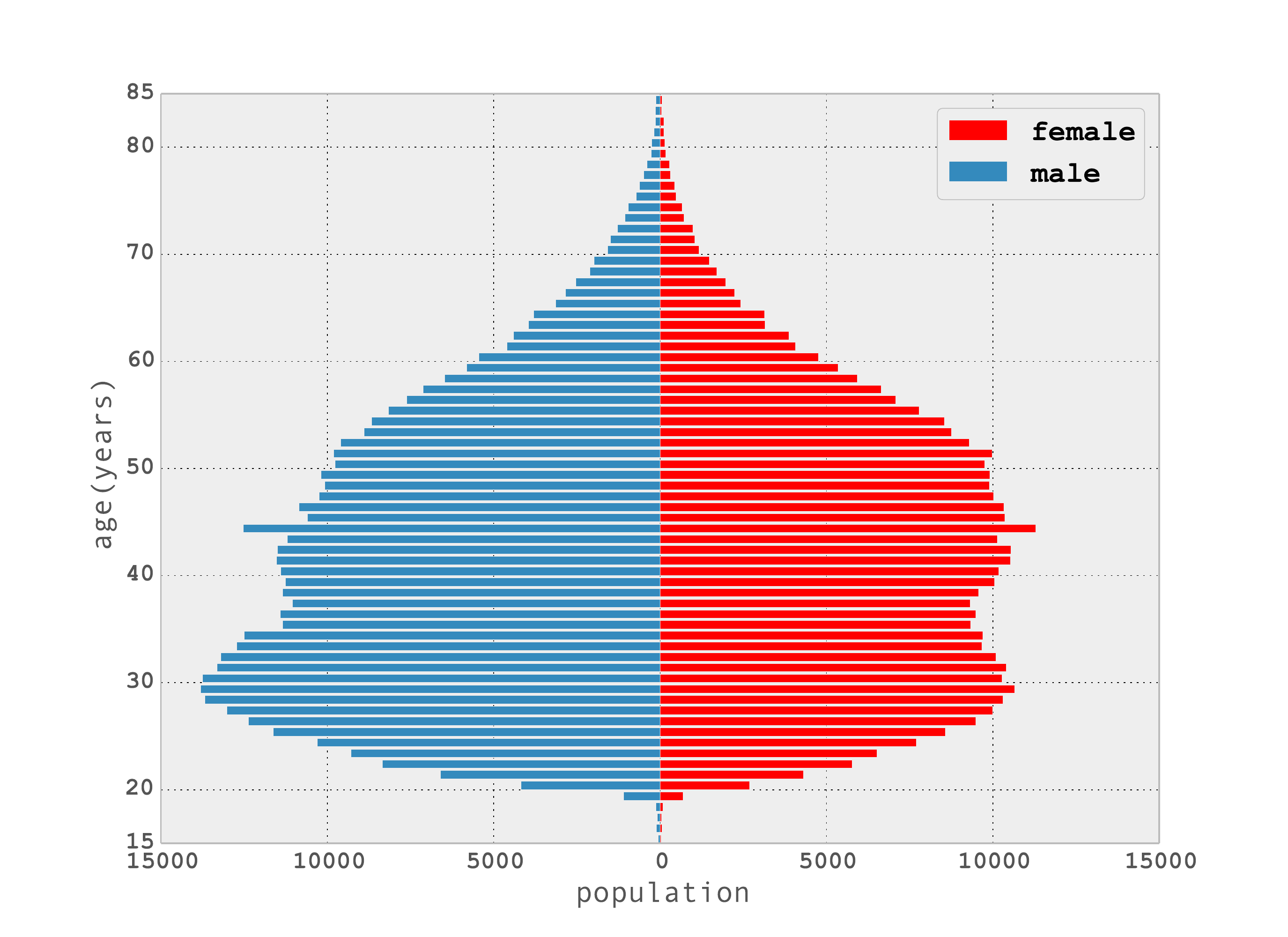}}
	\caption{Population pyramid.}
	\label{fig:age_histogram}
\end{figure}

In Fig.~\ref{fig:age_histogram} we plot the population pyramid of the seed nodes discriminating by gender.
We can see a bimodal structure for both genders with peeks centered around the ages of 30 and 45 years.
This double peak in the distribution is not observed in the Mexico age population census,
thus it is possibly indicative of some particular feature of the population of mobile phone
clients arising from special offers such as family plans.

The last step in our data preprocessing consisted in partitioning users ages into four age groups: below 25 years, from 
25 to 34 years, from 35 to 49 years and 50 years and above. The motivation for the chosen partition 
is grounded in marketing purposes of the mobile phone operator. The predictive algorithm will therefore 
infer one of these age groups for each of the nodes in $\calG$.

\begin{figure*}[ht]
\centering
\begin{minipage}{.495\textwidth}
	\centering
	{\includegraphics[trim=1.2cm 0.2cm 2.4cm 1.2cm, clip=true, width=1.0\linewidth]
	{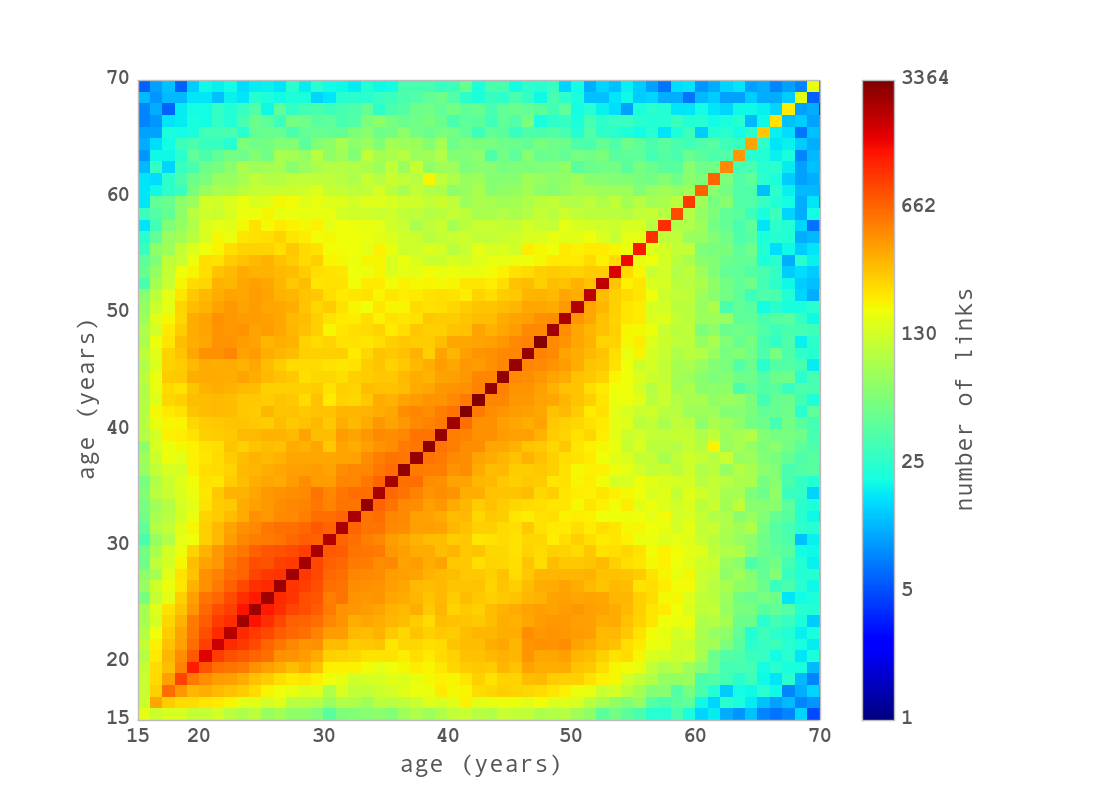}}
	(a) Communications matrix $C$.
\end{minipage}
\begin{minipage}{.495\textwidth}
	\centering
	{\includegraphics[trim=1.2cm 0.2cm 2.4cm 1.2cm, clip=true, width=1.0\linewidth]
	{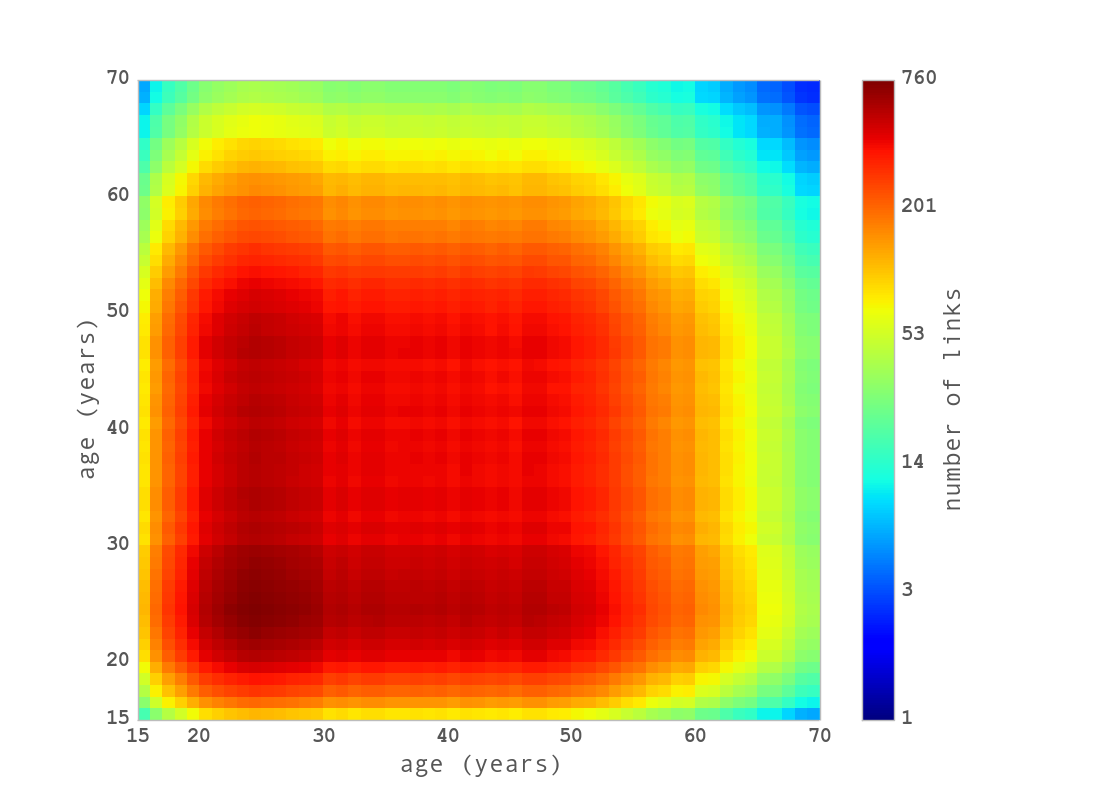}}
	(b) Random links matrix $R$.
\end{minipage}
\begin{minipage}{.495\textwidth}
	\centering
	{\includegraphics[trim=1.2cm 0.2cm 2.4cm 1.2cm, clip=true, width=1.0\linewidth]
	{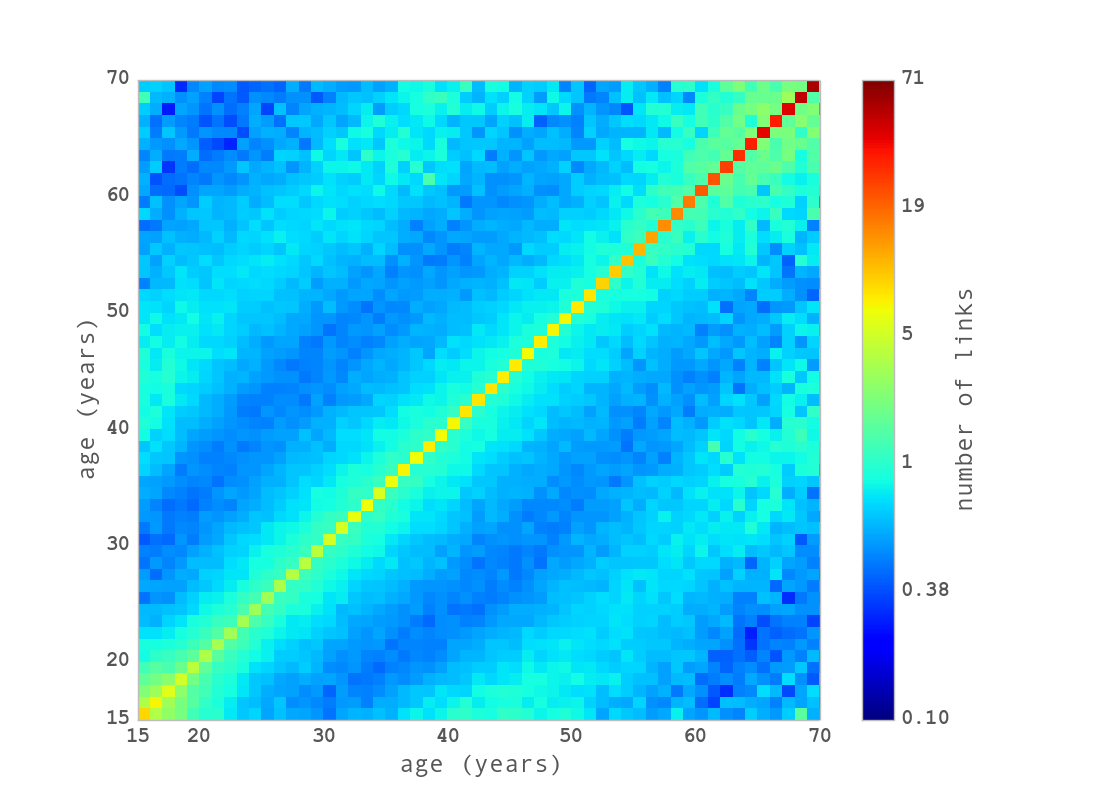}}
	(c) Difference between $C$ and $R$.
\end{minipage}
\begin{minipage}{.495\textwidth}
	\centering
	{\includegraphics[trim=1.2cm 0.2cm 2.4cm 1.2cm, clip=true, width=1.0\linewidth]
	{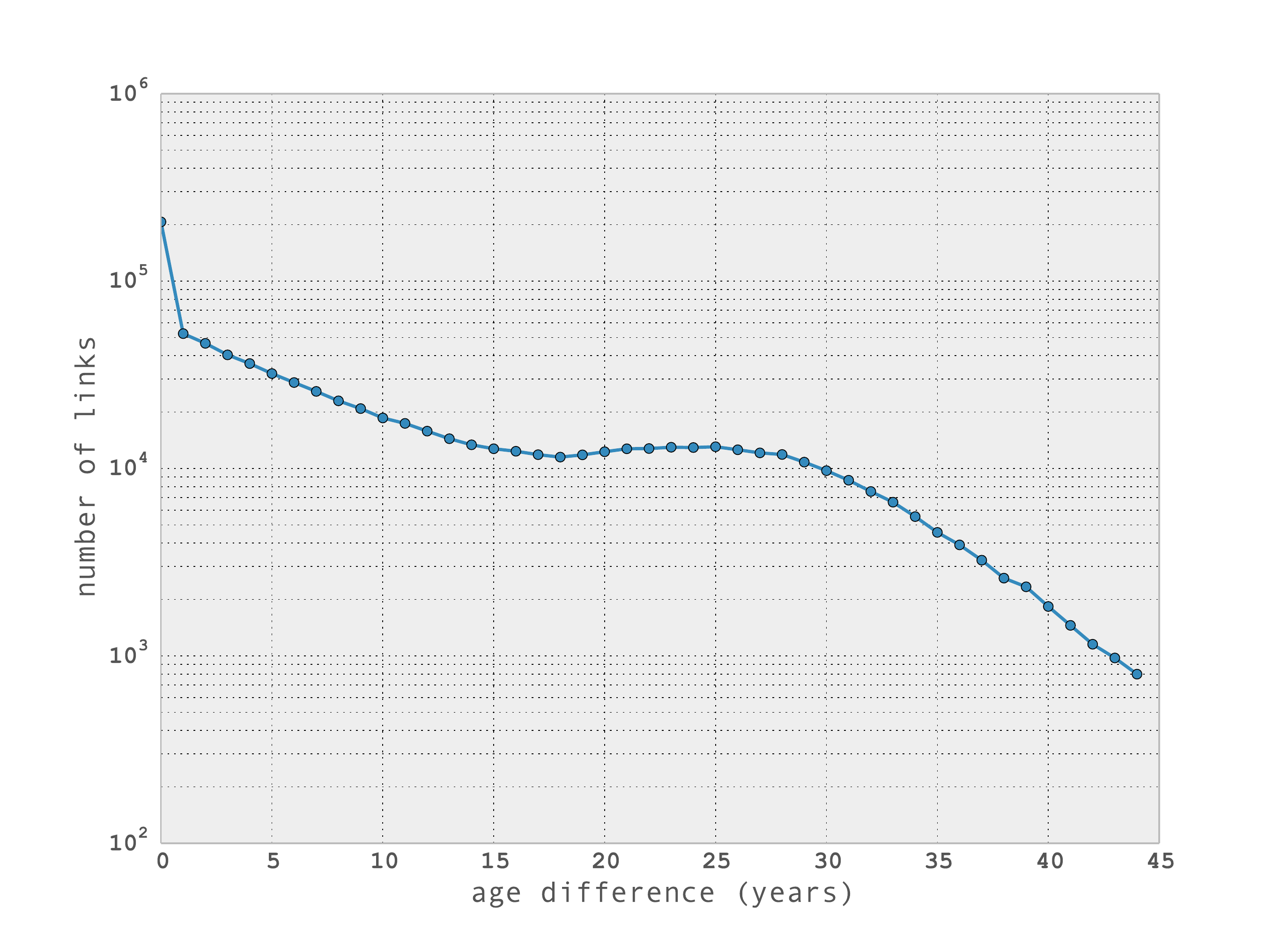}}
	(d) Number of links as function of age difference.
\end{minipage}
	\caption{Age homophily plots showing the communications matrix $C$, the random links matrix $R$, 
	the difference between $C$ and $R$, and the number of links according to the age difference.}
	\label{fig:communications}
\end{figure*}

\subsection{Age Homophily} \label{sec:age_homophily}

Homophily is defined as ``the principle that contacts between similar people occurs
at a higher rate than between dissimilar people'' \cite{mcpherson2001birds}.
This basic observation about human relations can be traced back to Aristotle, who
wrote that people ``love those who are like themselves'' \cite{aristotle1995complete}.
In particular, age homophily in social networks has been observed in the sociological literature,
for instance in the context of friendship in urban setting~\cite{fischer1977networks}
and other social structures~\cite{feld1982social}.
These historical studies were limited to hundreds of individuals, due to the difficulty
and cost of gathering data.
A recent study on the structure of the Facebook social graph~\cite{ugander5anatomy}
shows that a strong age homophily can be observed in a social graph with hundreds of millions
of users.
Our dataset allowed us to verify that the age homophily principle is valid
in the mobile phone communication network on a nation wide scale (the country of Mexico in our case).

As we mentioned in the introduction, 
graph based methods like the one we present in this work rely strongly on the ability of the 
graph topology to capture correlations between node attributes we are aiming 
to predict. A most fundamental structure is that of correlations between nodes and their 
neighborhoods. Figure~\ref{fig:communications}(a) shows the correlation matrix $\mtrx{C}$ 
where $\mtrx{C_{i,j}}$ is 
the number of links between users of age $i$ and age $j$ for the nodes in the the ground
truth $\calN_{GT}$. Though we can observe some smaller off diagonal peaks, we can see that it is 
most strongly peaked along the diagonal,
showing that users are much more likely to communicate with users of their same age. 
Performing a linear regression between the ages of linked users yields
a regression coefficient $r = 0.99$ and gradient $1.06$, confirming this observation.

To account for the population density bias shown in Fig.~\ref{fig:age_histogram}, 
we compute a surrogate correlation matrix $\mtrx{R}$ 
as the expected number of edges
between ages $i$ and $j$ under the null hypothesis of independence,
ploted in Fig.~\ref{fig:communications}(b):
\begin{equation}
\mtrx{R_{i,j}} = 
\frac{ | N_{GT}(i) | }{ | \calN_{GT} | } \times
\frac{ | N_{GT}(j) | }{ | \calN_{GT} | } 
\times | \calE_{GT} | 
\end{equation} 
where $N_{GT}(i) = \{ x \in \calN_{GT} : age(x) = i \}$ 
and $\calE_{GT} = \{ (x,y) \in \calE : x \in \calN_{GT} \wedge y \in \calN_{GT} \}$.
This matrix represents a graph with the same nodes as the original, but with random edges
(while maintaining the same number of edges as the original). 
Both $\mtrx{C}$ and $\mtrx{R}$ are represented with a logarithmic color scale.
If we subtract the logarithm of $\mtrx{R}$ to the logarithm of the original matrix $\mtrx{C}$, 
we can isolate the ``social effect'' (i.e. homophily) from the pure random connections, 
as can be seen in Fig.~\ref{fig:communications}(c).

Figure~\ref{fig:communications}(d) summarizes the number of links as a function of the age difference
$\delta$, showing a clear peak when the difference is $\delta = 0$.
The number of links decreases with the age difference, except around the 
value $\delta = 25$, where an interesting inflection point can be observed;
possibly relating to different generations (e.g. parents and children).
This phenomenon can also be seen in Fig.~\ref{fig:communications}(c), in 
the off diagonal bands at distance 25 years from the diagonal.

%% file: methods.tex
\section{Methods} \label{sec:methods}
The strong correlations shown in Fig.~\ref{fig:communications} indicate that 
the topological information of the network can be used to 
successfully infer the nodes age group (knowing the age group for a small subset of 
seed nodes). We describe below our algorithm that performs such task.

\subsection{Reaction-Diffusion Algorithm} \label{sec:reaction-diffusion-algorithm}

For each node $x$ in $\calG$ we define an initial state probability vector $\vect{g}_{x,0} \in \mathbb{R}^{C}$
representing the initial probability of the nodes age belonging to one of the $C = 4$ age categories 
defined in section~\ref{sec:dataset}. More precisely, each component of $\vect{g}_{x,0}$ is given by
\begin{equation}
\label{eq:inititalconditions}
	(\vect{g}_{x,0})_i = \begin{cases}
		\delta_{i,a(x)} & \text{if }  x \in \calN_{S}  \\
		1/C  & \text{if } x \not\in \calN_{S} 
	\end{cases}
\end{equation}
where $\delta_{i,a(x)}$ is the Kronecker delta function and $a(x)$ the age category
assigned each seed node $x$. For non seed nodes, equal probabilities are assigned to 
each category. These probability vectors are the set of initial conditions for the algorithm.

The evolution equations for the probability vectors $\vect{g}$ are then as follows
\begin{equation}
\label{eq:difusiongeneral_con_pesos}
	\begin{split}
		\vect{g}_{x,t} &= (1-\lambda) \; \vect{g}_{x,0}+ \lambda \; \frac{\sum_{x\sim y} w_{y,x} \, \vect{g}_{y,t-1} }{\sum_{x\sim y} w_{y,x}} 
	\end{split}
\end{equation} 
where $x \sim y$ is the set of $x$'s neighbours and $w_{y,x}$ is the weight of the link 
between nodes $x$ and $y$.
For our purposes all link weights are set to one, thus the second term simplifies
to a normalized average over the values of the neighborhood nodes for $x$:
\begin{equation}
\label{eq:difusiongeneral}
		\vect{g}_{x,t} = (1-\lambda) \; \vect{g}_{x,0}+ \lambda \; \frac{\sum_{x\sim y} \vect{g}_{y,t-1} }{|\{y:x\sim y\}|} .
\end{equation}
The evolution equation is made up of two distinctive terms: the first term can be thought of 
as a memory term where in each iteration the node remembers its initial state $\vect{g}_{x,0}$, while the second term is a mean 
field term of the average values of the nodes neighborhood
for $x$.
The model parameter $\lambda \in [0,1]$ defines the relative importance of each of these terms. 
The influence of $\lambda$ on the performance of the algorithm will be analyzed in
section~\ref{sec:model-parameters}.

We note that the evolution of each term in the probability vectors are totally decoupled from one another. This last assumption in the algorithm relies on the strong homophily suggested by the correlation plots of Fig.~\ref{fig:communications}.

Next, we briefly outline our motivation for calling it 
a \emph{reaction-diffusion} algorithm.
A reaction–diffusion model explains how the concentration of one or more (typically chemical) substances changes under the influence of two processes: local reactions and diffusion 
across the medium ~\cite{Nicolis1977}.

Let $\mtrx{A}\in \mathbb{R}^{| \calN | \times | \calN |}$ be the adjacency matrix of $\calG$,
and $\mtrx{D}\in \mathbb{R}^{| \calN | \times | \calN |}$ the diagonal matrix with the degree of each node in $\calG$,
that is $\mtrx{A_{i,j}} = w_{i,j}$ and $\mtrx{D_{i,i}} = deg(i)$. Furthermore 
let $\mtrx{L} \in \mathbb{R}^{| \calN | \times | \calN |}$ be the matrix defined as $\mtrx{L} = \mtrx{D} - \mtrx{A}$. 
We can now rewrite equation~\eqref{eq:difusiongeneral} as
\begin{equation}
	\vect{g}_{t}^{a} = (1-\lambda) \vect{g}_{0}^{a} + \lambda 
	\mtrx{D}^{-1/2} [ \mtrx{D} - \mtrx{L}] \mtrx{D}^{-1/2} \vect{g}_{t-1}^{a}
\end{equation}   
where $\vect{g}^{a}\in\mathbb{R}^{|\calN|}$  now runs over all nodes, for each age probability category $a$. We are able 
to do this since from equation~\eqref{eq:difusiongeneral}, the probabilities of each category evolve independently.
We now rewrite the above equation using the graph Laplacian for $\calG$ defined as $\mtrx{\calL}$ = $\mtrx{D}^{-1/2} \mtrx{L} \mtrx{D}^{-1/2}$~\cite{Chung1997}:
\begin{equation}
	\vect{g}_{t}^{a} = (1-\lambda) \vect{g}_{0}^{a} - \lambda \mtrx{\calL} \vect{g}_{t-1}^{a} + \lambda \vect{g}_{t-1}^{a}
\end{equation}
finally reordering terms we get
\begin{equation}\label{eq:laplacian}
	\vect{g}_{t}^{a}-\vect{g}_{t-1}^{a} = (1-\lambda)(\vect{g}_{0}^{a}-\vect{g}_{t-1}^{a}) 
	- \lambda \mtrx{\calL} \vect{g}_{t-1}^{a}
\end{equation}
where on the left hand side we have the discrete derivative of $\vect{g}^{a}$ with respect to time, the first term on the 
right, $(1-\lambda)(\vect{g}_{0}^{a}-\vect{g}_{t-1}^{a})$, is the reactive term which drives $\vect{g}_{t}^{a}$ towards $\vect{g}_{0}^{a}$ and the second term $\lambda \mtrx{\calL} \vect{g}_{t-1}^{a}$ can be thought of as a diffusive 
term where the Laplacian operator in continuous space has been replaced by the Laplacian of the graph $\calG$ over which diffusion is taking place. From equation~\eqref{eq:laplacian}, we can see that $\lambda$ tunes the relative importance of the seed nodes information $\vect{g}_{0}$ and the topological properties of $\calG$ given by the Laplacian $\mtrx{\calL}$. We have therefore gone from a local description of the model in equation~\eqref{eq:difusiongeneral_con_pesos}, to a global description in equation~\eqref{eq:laplacian}.

\subsection{Topological Metrics} \label{sec:topological-metrics}

One of the main goals of this work is to uncover topological properties 
for a given node that increase its chances of being correctly predicted, 
and characterize (based on these properties)
a subset of nodes in $\calG$ for which our algorithm works particularly well.
For this, we describe three topological metrics used to characterize each node in $\calG$. 
The first
metric is the number of seeds in a node's neighborhood which we denote \emph{SIN} (seeds in neighborhood).
We expect \emph{SIN} to be a relevant metric given the strong correlations shown in Fig.~\ref{fig:communications}.  
The second metric is the topological distance of a node to the seed set which 
we denote \emph{DTS} (distance to seeds), 
where distance is defined as the shortest path from the node to its closest seed. 
The third
metric is the degree $d$ of a node in the network, that is, the size of its neighborhood.
Nodes with high degree are more likely to be strongly connected to the seed nodes, therefore we expect their response to the
reaction diffusion algorithm, all else equal, to be significantly different than that of a node with low degree.

%% file: results.tex
\section{Results} \label{sec:results}

In this section we first present the results for the predictive power
of the reaction-diffusion algorithm over the whole network $\calG$. To construct
the mobile phone network we start with a directed edge list consisting
of $252,248,440$ edges where an edge can represent outgoing communications
from client $x$ to user $y$ and incoming communications from user $y$ to client
$x$ distinctively. This set of edges involves $93,797,349$ users.
After pruning users with high degree (above 100), or not belonging to any connected 
component containing seed nodes, we end up with a 
resultant graph $\calG_T$ with  
$|\calE_T| = 125,433,585$ edges (where edges have been symmetrized) and 
$|\calN_T| = 71,949,815$ nodes, of which $|\calN_S| = 493,871$ are
seed nodes and $|\calN_V| = 164,768$ are validation nodes for which we know
their actual age group. These are used to test the inference power of our algorithm.    

As stated in equation~\eqref{eq:inititalconditions}, 
the initial conditions for each node's probability
vector $\vect{g}_{x,0}$ is that of total certainty for seed nodes and no information
for the remaining nodes in $\calG_T$. At each iteration, each
$\vect{g}_{x,t}$ in~\eqref{eq:difusiongeneral} updates its state
as a result of its initial state and the mean field resulting from the probability
vector of its neighbors. To reduce diffusion noise, in every time step
we consider the mean field generated only by the neighbors of $x$  
that have received some information from the seed nodes.
After the last iteration $t_{\textrm{end}}$ we assign the age group with highest probability in $\vect{g}_{x,t_{\textrm{end}}}$
to each node in $\calG_T$. Unless stated otherwise, we set the value of $\lambda=0.5$ and the number of 
iterations to $30$ (which is $10$ iterations more than the maximum distance from a node to the seed set: $\mbox{\emph{DTS}}_{\textrm{max}}=20$).

\subsection{Performance by Age Group}

In Table~\ref{tb:demographicsfull} we present the results for the performance of the
reaction diffusion algorithm discriminating by age group.
The age group
demographics predicted by the algorithm, both for the whole graph $\calN_T$ and
the validation set $\calN_V$ are similar to that of the seed set $\calN_S$, that is,
the algorithm does a good job at preserving the age group demographics
of the seed set for all nodes in $\calG$. The last column shows the fraction
of correctly validated nodes for each age group ($hits$) we refer to as the performance
or inference power of the algorithm. The
age group between 35 and 50 years is significantly better predicted than the other
three age groups with a performance of 52.3\%. 

\begin{table}[ht]
\centering
\begin{tabular}{  l  r  r  r  r }
\toprule
Age group & $\%\calN_T$ & $\%\calN_S$ & $\%\calN_V$ & \%\emph{hits}\\
\midrule
<25 & $5.4\%$ &  $8.6\%$ & $4.6\%$ & $26.6\%$\\
25-34 & $32.7\%$ &  $27.7\%$ & $32.3\%$ & $44.6\%$\\
35-50 & $37.6\%$ &  $35.9\%$ & $35.8\%$ & $52.3\%$\\
50+ & $24.4\%$ &  $27.8\%$ & $27.3\%$ & $45.0\%$\\
\bottomrule
\end{tabular}
\caption{Performance by age group.}
\label{tb:demographicsfull}
\end{table}

A possible  
explanation for the increased performance for the age group 35-50 can be found by
looking at the correlation plot in Fig.~\ref{fig:communications} where we observe
not only strong correlations and high population densities for this age group,
but it is also less fooled by the offside peaks mostly concentrated within the
age group 25-34. Again we see the importance of the correlation
structure among neighboring nodes. Finally, the overall performance for the
entire validation set was $46.6\%$ and we note that a performance based on random
guessing without prior information would result in an expected performance of $25\%$,
or an expected performance of $\sim 36\%$ if we set all nodes age goup to the most 
probable category ($35-50$).

\subsection{Performance Based on Topological Metrics}\label{sec:performance-metrics}

We next focus our attention on how our algorithm performs for
different  subsets of $\calG_T$ selected according to the  
topological metrics described in section~\ref{sec:topological-metrics}. The main
purpose of this section is to show that, given a set of seed
nodes, there are specific topological features characterizing
each node in the network that allows us to, a priori, select a set 
of nodes with optimal expected performance. We label each node
in our validation set with the three topological metrics described 
in section~\ref{sec:topological-metrics}. 

\begin{figure}[t]
    \centering
    {\includegraphics[trim=1.5cm 0cm 1.5cm 1.0cm, clip=true, width=0.95\linewidth]
    {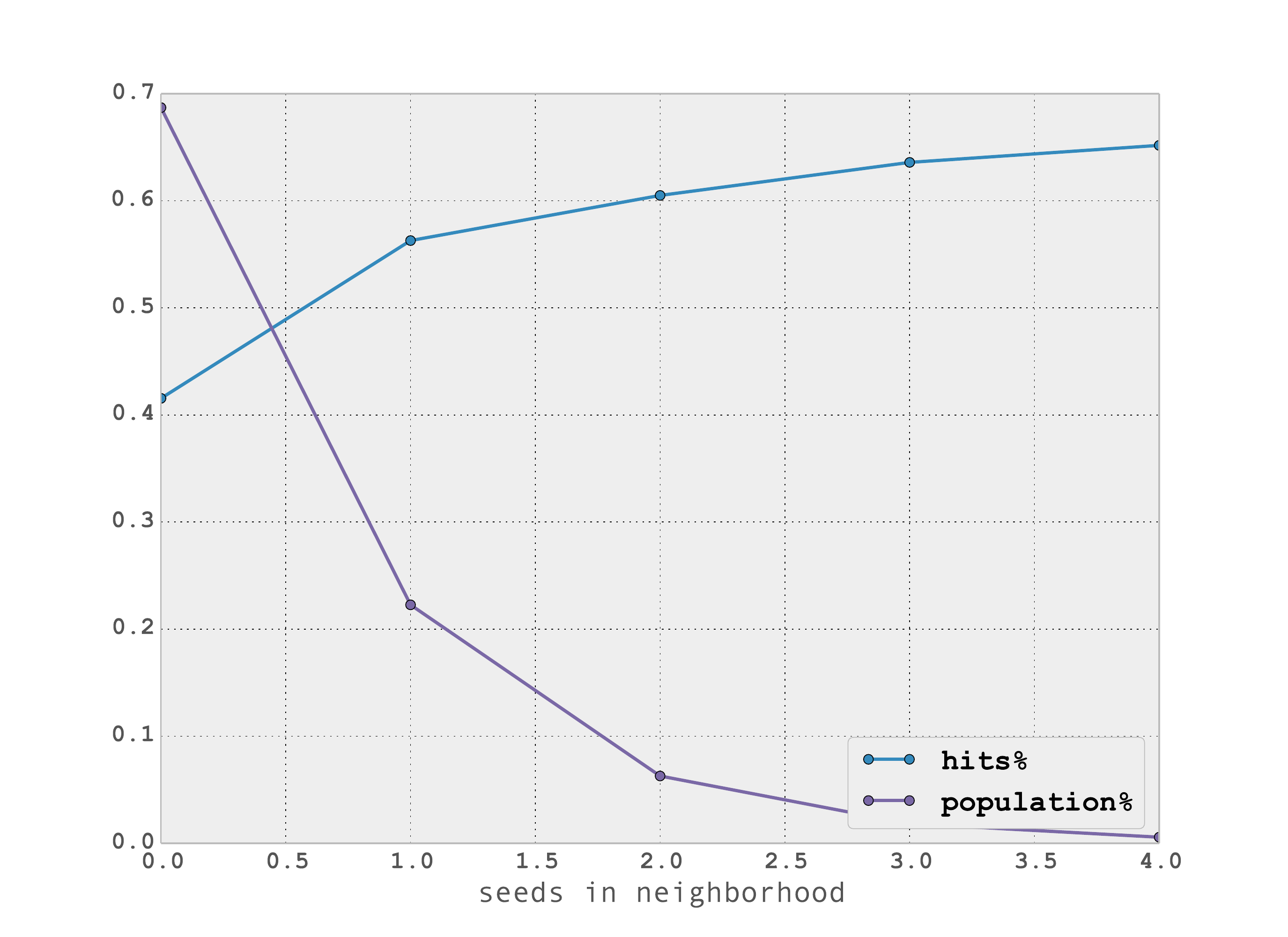}}
    \caption{Performance and population as function of \emph{SIN} (seeds in network).}
    \label{fig:seedsinego}
\end{figure}

Figure~\ref{fig:seedsinego} plots the performance (\emph{hits}) of
the algorithm over a set of nodes as a function of \emph{SIN} (number
of seeds in the nodes neighborhood). The algorithm performs
worst for nodes with no seeds in the immediate neighborhood with \emph{hits} = 41.5\%,
steadily rising as the amount of seeds increase
with a performance of \emph{hits} = 66\% for nodes with 4 seeds in their neighborhood.
We also see that the amount of nodes decreases exponentially
with the amount of seeds in their neighborhood. An interesting feature
we observed is that it is the total amount of seed nodes and not the proportion
of seed nodes that correlate most with the performance of the algorithm (this last
observation is not shown in the figure).

\begin{figure}[t]
    \centering
    {\includegraphics[trim=1.5cm 0cm 1.5cm 1.0cm, clip=true, width=0.95\linewidth]
    {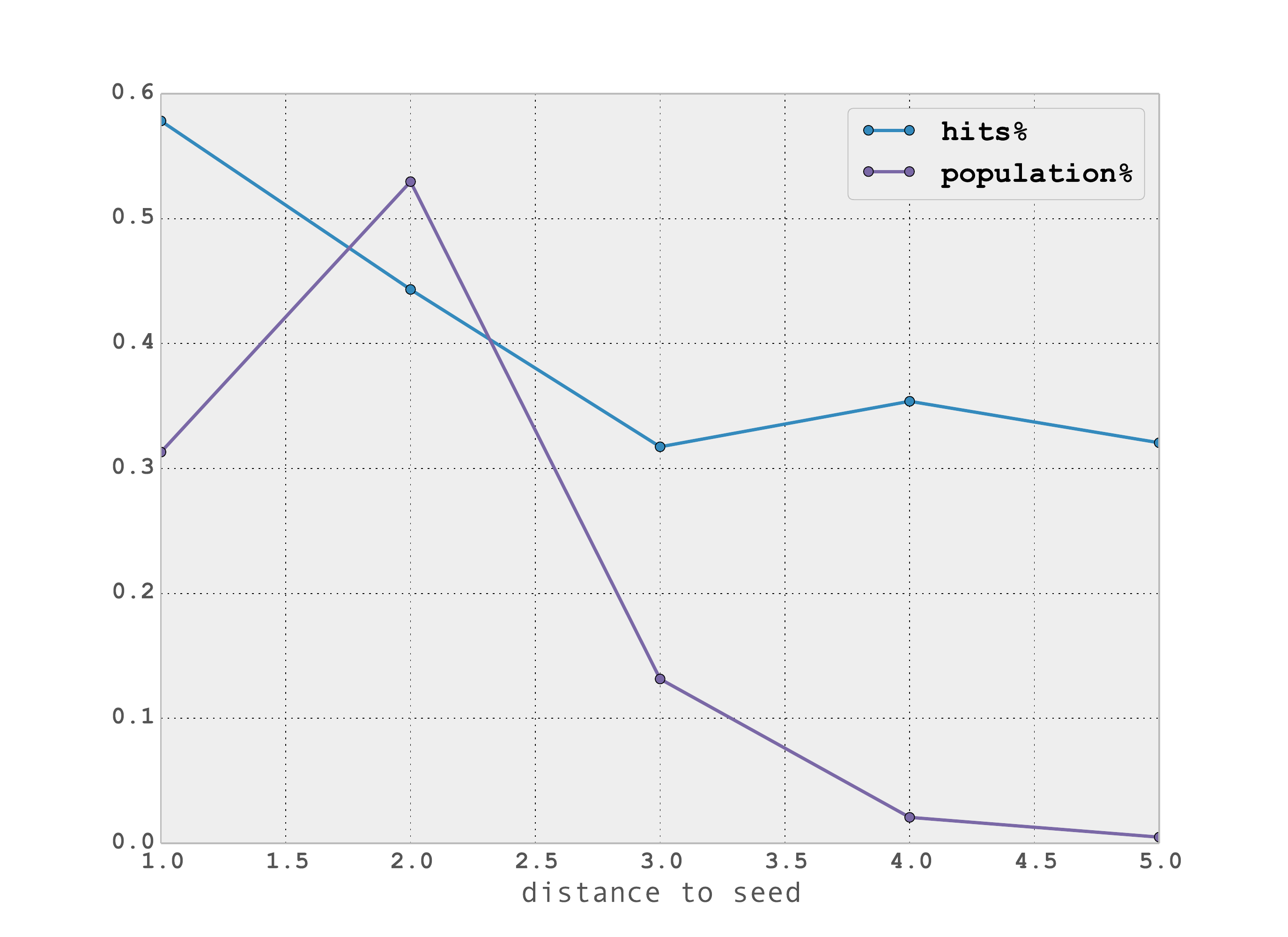}}
    \caption{Performance and population as function of \emph{DTS} (distance to seeds set).}
    \label{fig:distancetoseeds}
\end{figure}

We next examine how the algorithm performs
for nodes in $\calG_T$ that are at a given distance to the seed set (\emph{DTS}).
In Fig.~\ref{fig:distancetoseeds} we plot the population size
of nodes as a function of their \emph{DTS}. The most frequent
distance to the nearest seed is 2, and almost all nodes
are at distance less than 4. This implies
that after four iterations of the algorithm, the seeds information
have spread to most of the nodes in $\calG_T$. This figure also shows
that the performance of the algorithm decreases as the distance of a node
to $\calN_S$ increases.

Both \emph{SIN} and \emph{DTS}, although topological quantities themselves, are defined in terms of
the distribution of the seeds set, and are not intrinsic properties the graph $\calG_T$. 
We therefore look at a basic intrinsic topological property of a node
$x$ in $\calG_T$, namely its degree $d(x)$.

\begin{figure}[t]
	\centering
    {\includegraphics[trim=1.7cm 0cm 2.0cm 1.0cm, clip=true, width=0.95\linewidth]
	{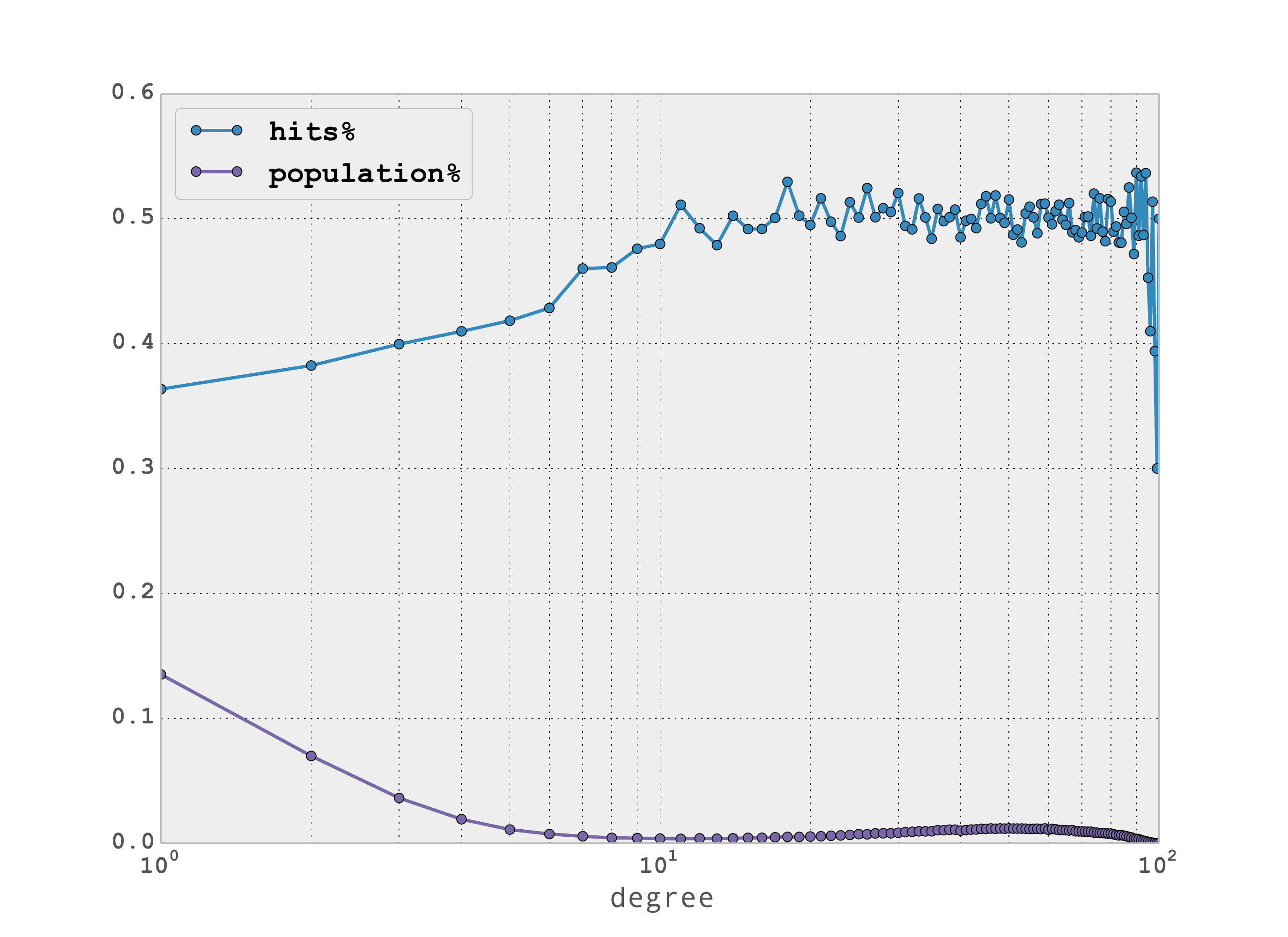}}
	\caption{Performance and population as function of the nodes degree.}
	\label{fig:nodesdegree}
\end{figure}

In Fig.~\ref{fig:nodesdegree} we see that the performance
of the algorithm is lowest for nodes with small degree and gradually increases as
the degree increases reaching a plateau for nodes with $d(x) > 10$. 

Understanding what makes degree
values close to 10 an inflection region in the performance of the reaction-diffusion
algorithm is something we will address in a future work, but it is here worth remarking
that $\calG_T$ is an essentially incomplete graph, where we are missing all the communication
links among users that are not clients of the mobile operator. An important topological
consequence of this is that whereas the average degree over the whole graph $\calG_T$
is $3.48$, the average degree for client nodes is $37.53$.

\begin{table}[ht]
\center
\begin{tabular}{@{}lccc@{}} \toprule
& \multicolumn{3}{c}{Distance to Seeds} \\ \cmidrule(l){2-4}
Degree & 1 & 2 & 3 \\  \midrule
{[}1, 2] & 0.415057 & \textbf{0.465881} & 0.313735 \\ 
(2, 29] & 0.581540 & 0.445736 & 0.334968 \\ 
(29, 48] & \textbf{0.618067} & 0.435747 & 0.299578 \\ 
(48, 66] & 0.588592 & 0.438933 & 0.288136 \\ 
(66, 100] & 0.566038 & 0.441583 & 0.206897 \\ \bottomrule
\end{tabular}
\caption{Performance as function of degree and distance to seeds.}
\label{tb:optimalmetrics}
\end{table}

Having shown the strong correlations between the three topological
metrics and the algorithm's performance -- which allows us to select
optimal values for each metric independently -- we next examine whether
we can find an optimal combination of these metrics.
Table~\ref{tb:optimalmetrics} shows the performance of our algorithm for different values 
of their \emph{DTS} and their degree. 
Even though we have shown that
performance increases with smaller distance to seeds and higher degree, here we find
that this rule can be violated for specific ranges of the nodes degree -- as is shown
for nodes with degree between 1 and 2
where performance is optimal for nodes at distance 2 from the seed set. 
Most interestingly, we find that selecting a subset based
only on the nodes degree and distance to seeds, the optimal set is at distance 1
from the seed set and with degree between 29 and 48 for which
the performance  of the algorithm increases to \emph{hits} = 61.8\% with 20,050 nodes in $\calN_V$
satisfying these set of conditions. This last result indicates that it is not sufficient
to look at each metric independently to select the subset of nodes for optimal performance
of the algorithm. We also looked at the performance of the algorithms as a function of \emph{SIN} and 
degree, and also found subsets of nodes for whom performance increased to similar values 
(table not shown for space reasons).

\subsection{Preserving Age Demographics} \label{sec:pyramidscaling}
A salient feature of our algorithm is that the demographic information being spread 
is not the age group itself, but a probability vector for each age group.
In each iteration, the algorithm does not collapse the information in each node
to a preferred value; instead it allows the system to evolve as a probability state over the network, 
and only at the end of the simulation, when we ``observe'' the system, are we required to collapse
the probability vectors to specific age groups. 
Choosing how to perform this collapse is not an obvious matter,
in the previous sections we selected for each node the category with highest probability.
This is a natural approach, 
and as we have shown, the algorithm's performance is satisfactory. 
A draw back of this approach is that,
when collapsing the probability vector at each node, we ignore the resultant probabilities 
for all other nodes in the network which could inform us in making a better selection choice.
We might want to collapse the probability state of the system as a whole and not for each node independently. 
For instance, if we want to impose external constraints on our solution, 
namely that the age group distribution for the whole network be that of the seed set, 
we can do this optimally using the following algorithm,
which we call Population Pyramid Scaling (PPS).

\begin{algorithm}
\ForEach{\emph{node $i$ and group $a$}}{
  compute the probability $p_{i,a}$ that node $i$ belongs to group $a$ 
  using our (unconstrained) algorithm\;
}
Create an ordered list $T$ of tuples $(i, a, p_{i,a})$\; 

Sort list $T$ in descending order by the column $p_{i,a}$.
The list $T$ will be iterated starting with the element with the highest probability\;

\ForEach{\emph{element $(i, a, p_{i,a}) \in T$}}{
		\If{\emph{node $i$ has not been assigned to a group}}{
		  \If{\emph{less than $N_a$ nodes assigned to group $a$}}{
		      assign node $i$ to group $a$\;
		  }
		}
}
\caption{Population Pyramid Scaling}
\label{algo:pps}
\end{algorithm}

Note that the population to predict has size $| \calN_T |$.
For each age group $a$ we compute the number of nodes $N_a$ 
that should be allocated to group $a$ in order to 
satisfy the distribution constraint (the age distribution of $\calN_{GT}$),
and such that $\sum_{a = 1}^{C} N_a = | \calN_T |$ 
(where $C$ is the number of age groups).
The PPS procedure is described in Algorithm~\ref{algo:pps}.

The performance of our algorithm constraining the demographic population of $\calG_T$ to that of the seed nodes $\calN_S$
is $46.5\%$, almost identical to $46.6\%$ obtained without PPS. The reason for this is that the algorithm (without PPS) was able to preserve a distribution over the whole graph very similar to that of 
the seed nodes as we showed in Table~\ref{tb:demographicsfull}.
Therefore in this case, the demographic constraint has not significantly changed our results.

\subsection{Analysis of the Clients Subgraph}

In section~\ref{sec:performance-metrics} we mentioned that the graph $\calG_T$ has
two distinct types of nodes, client and non client nodes with considerably
different average degree. Motivated by this difference, and also
considering the distinctive importance from a business point of view of understanding
the predictive power of our algorithm on the client nodes, we performed
the same study as above for the graph $\calG_C$ consisting only of client nodes.
This graph is made up of $| \calE_C | = 6,484,571$ edges 
and $| \calN_C | = 3,225,538 $ nodes of which 349,542 nodes  
are seed nodes and 143,240 nodes are validation nodes. Note the average degree
of the client nodes has droped from $37.53$ in $\calG_T$ to 4.02 in $\calG_C$.

\begin{table}[ht]
\center
\begin{tabular}{  l  r  r  r  r }
\toprule
Age& $\%\calN_T$ & $\%\calN_S$ & $\%\calN_V$ & \%\emph{hits}\\
\midrule
<25 & $5.3\%$ &  $7.1\%$ & $4.6\%$ & $27.6\%$\\
25-34 & $28.7\%$ &  $29.9\%$ & $28.4\%$ & $50.9\%$\\
35-50 & $41.7\%$ &  $34.6\%$ & $37.2\%$ & $53.4\%$\\
50+ & $24.3\%$ &  $28.4\%$ & $29.8\%$ & $48.0\%$\\
\bottomrule
\end{tabular}
\caption{Performance by age group on the clients subgraph $\calG_C$.}
\label{tb:demographicsclientes}
\end{table}

In Table~\ref{tb:demographicsclientes}
we summarize the performance of the reaction-diffusion algorithm discriminated by age group.
As for the full graph $\calG_T$, the algorithm did a good job in preserving the
age group demographics of the seed nodes over all the nodes in $\calG_C$, in particular
for the validation nodes. The last column indicates the performance of the algorithm 
for each age group, where two observations are in order: first we notice that the overall
performance of $49.9\%$ is significantly better than the performance of $46.6\%$ for
the full graph $\calG_T$. Second the relative performance of each age group is quite
similar for both graphs, plausibly indicating that it is mainly governed by the
correlation structure described in Fig.~\ref{fig:communications}. 

\begin{figure}[t]
    \centering
    {\includegraphics[trim=1.5cm 0cm 1.5cm 1.0cm, clip=true, width=0.95\linewidth]
    {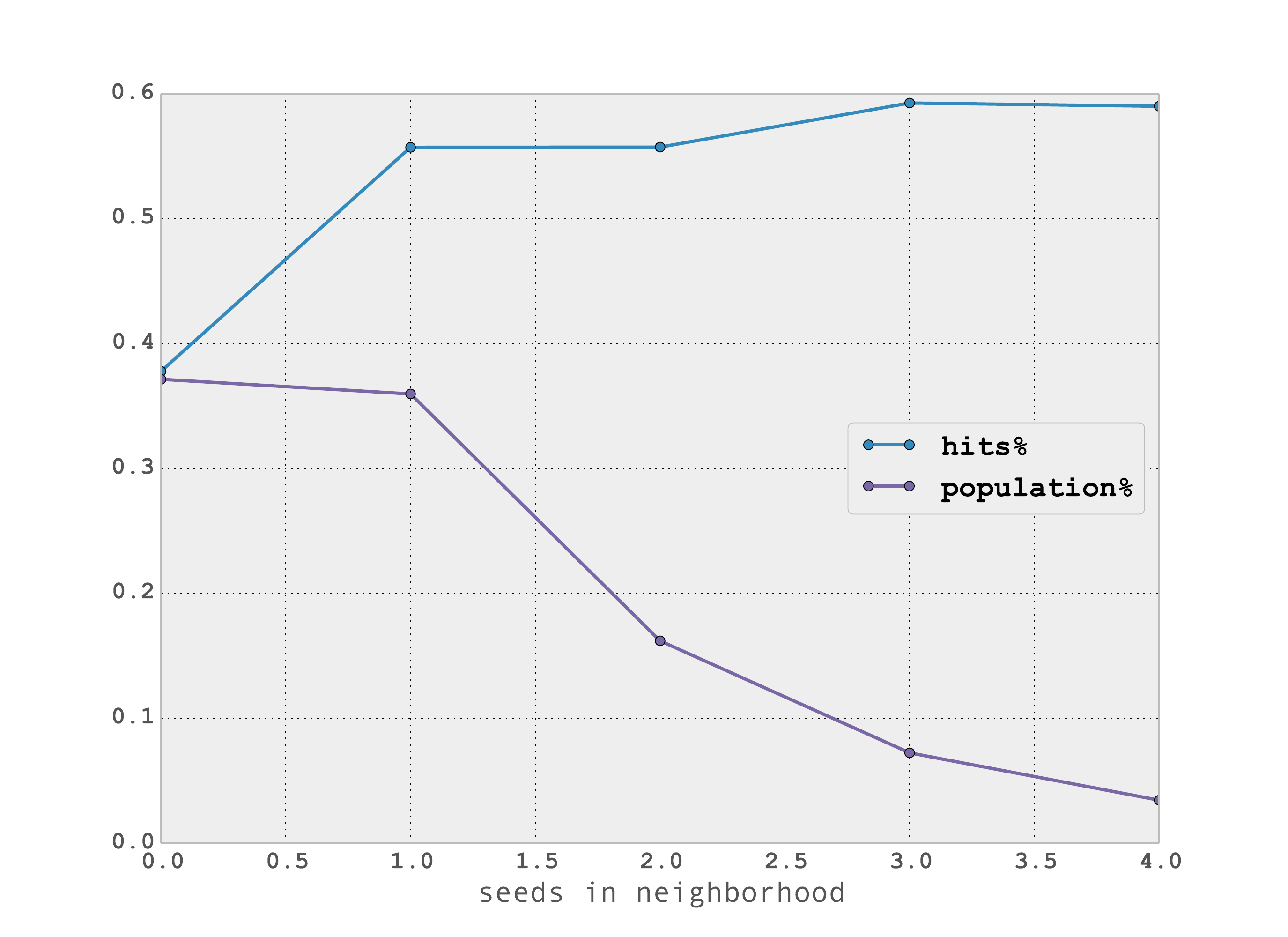}}
    \caption{Performance as function of \emph{SIN} (seeds in egonetwork) on the clients subgraph $\calG_C$.}
    \label{fig:seedsinegoclientes}
\end{figure}

\begin{figure}[t]
    \centering
    {\includegraphics[trim=1.5cm 0cm 1.5cm 1.0cm, clip=true, width=0.95\linewidth]
    {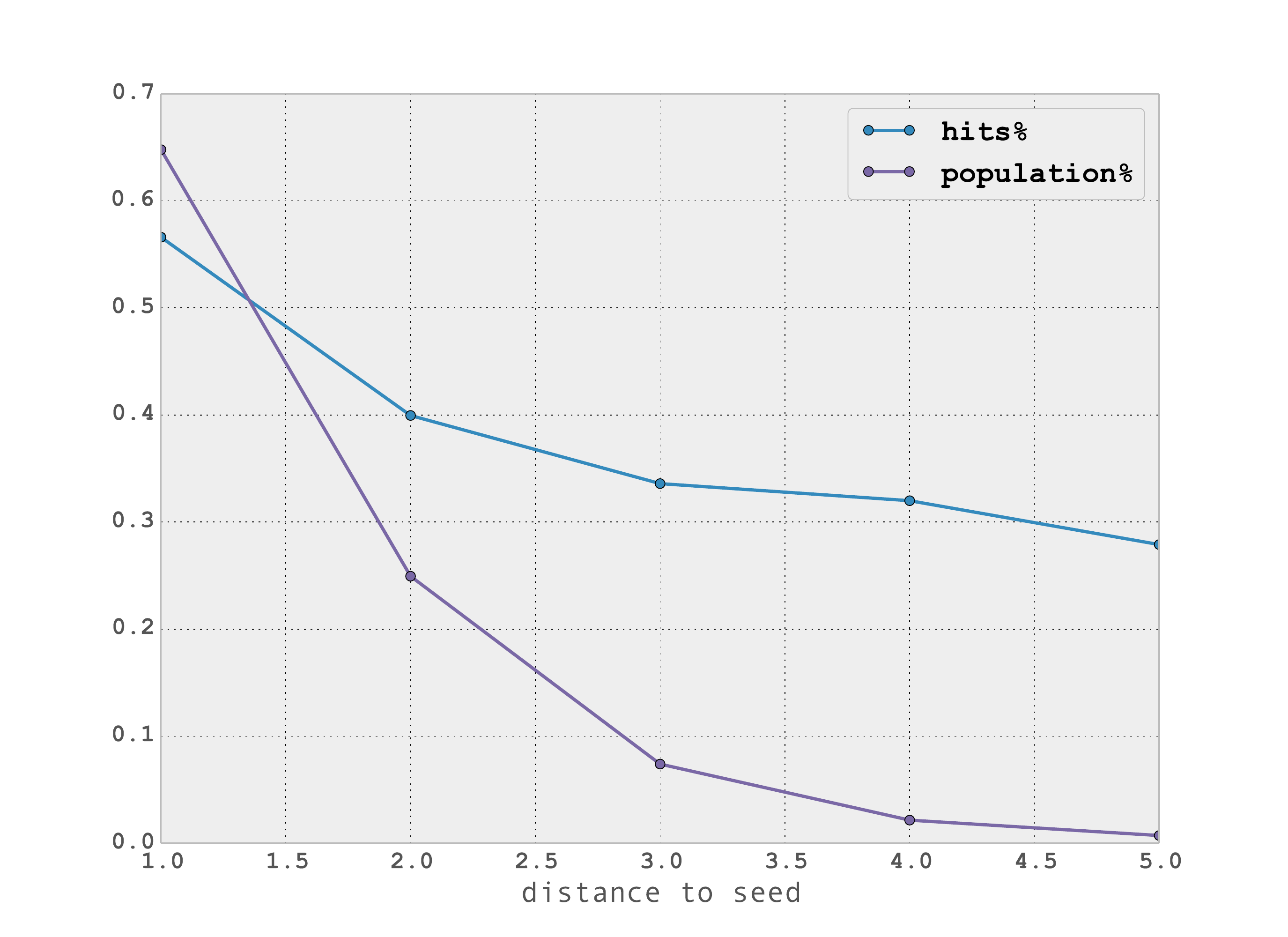}}
    \caption{Performance as function of \emph{DTS} (distance to seeds) on the clients subgraph $\calG_C$.}
    \label{fig:distancetoseedsclientes}
\end{figure}

Following the same analysis done for the full graph $\calG_T$ we looked
at how the performance of the reaction-diffusion algorithm correlates to
the three topological metrics. These results are shown in Figures~\ref{fig:seedsinegoclientes}
and \ref{fig:distancetoseedsclientes}.
Comparing these
plots with those for graph $\calG_T$ we can notice several topological differences between
them. First we notice that while in $\calG_T$ only $\sim 30\%$ of the validation nodes have
at least one seed in their neighborhood, over $65\%$ of nodes in $\calG_C$ have at least
one seed in their neighborhood. This is not surprising given that both validation and seed nodes
are client nodes. Most interestingly, even though for both graphs the performance of the algorithm
increases as their \emph{SIN} increases, for a fixed number of seeds the
algorithm does better in the full graph than in the client graph. This is quite surprising
given that the overall performance over the full graph is worse than in the client graph.

In Fig.~\ref{fig:distancetoseedsclientes} we see 
that \emph{DTS} for client nodes peaks at \emph{DTS} = 1, falling 
exponentially as \emph{DTS} increases (whereas in the full graph the peak was observed at \emph{DTS} = 2).
Performance of the algorithm again correlates 
strongly with distance to the seed set.

Performance as function of the degree increased as in the study for the whole 
graph $\calG_T$, though less abruptly (figure not shown for space reasons).

\subsection{Sensitivity to Model Parameters} \label{sec:model-parameters}
So far we have analyzed the performance of the reaction-diffusion algorithm
without addressing the issue of model parameters selection, namely the 
parameter $\lambda$ presented in section~\ref{sec:reaction-diffusion-algorithm}
-- which tunes the relative importance of the reaction and diffusion terms --
together with the number of iterations 
of the algorithm, which determines to which extent the age information in the seed set
spreads across the network. In this section we examine these two parameters for the client network $\calG_C$. 

\paragraph{Parameter $\lambda$}

For $\lambda=0.0$ (no diffusion) the performance is $28.1\%$ but as soon  
as the diffusion term is slightly turned on, $\lambda=10^{-7}$, the performance
of the algorithm jumps to $48.3\%$ and remains almost constant for $0 < \lambda < 1$ and drops suddenly
to a performance of $42.5\%$ for $\lambda=1.0$ (total diffusion with no reactive term).
We conclude from these results two things. First, the reactive term is important, increasing the performance from $42.5\%$ (no reactive
term) to $48.8\%$ (optimally choosing the reactive term). Second, inside the boundary values
for $\lambda$ the performance remains almost unchanged and thus the algorithm is very robust to parameter
perturbations. In other words, our algorithm needs some diffusion and reactive terms but the relative
weights of these are unimportant.

\paragraph{Convergence} 

An important property of the model we address here is that of the model's time scale, 
that is, how long it takes the system to reach a stationary state. 
Stationarity analysis in complex networks such as the one we study here is a hard problem, thus for the purpose of this work we limit ourselves to the notion of performance.
We therefore say the algorithm reaches a stationary state after $t$ iterations 
if its performance remains unchanged for further iterations. 
We have seen that most nodes in the graph are not directly connected to a seed node,
and that the diffusion term in Equation~\eqref{eq:difusiongeneral} spreads from neighbor to neighbor
one iteration at a time.
We therefore expect that as iterations increase, the seed nodes'
information will spread further across the network. 
Our experiments show that after
a single iteration, the performance of the reaction-diffusion algorithm is already at $45.5\%$ gradually increasing
its performance to a stationary value of $\sim 50\%$ after 5 iterations.
As conclusion, setting the number of iterations as $t_{\textrm{end}} = 30$ gives us a good margin to ensure convergence
of the algorithm.

\subsection{Optimal Node Subset using Probability Vector Information}

\begin{figure}[t]
	\centering
    {\includegraphics[trim=1.0cm 0cm 0cm 1.0cm, clip=true, width=0.95\linewidth]
	{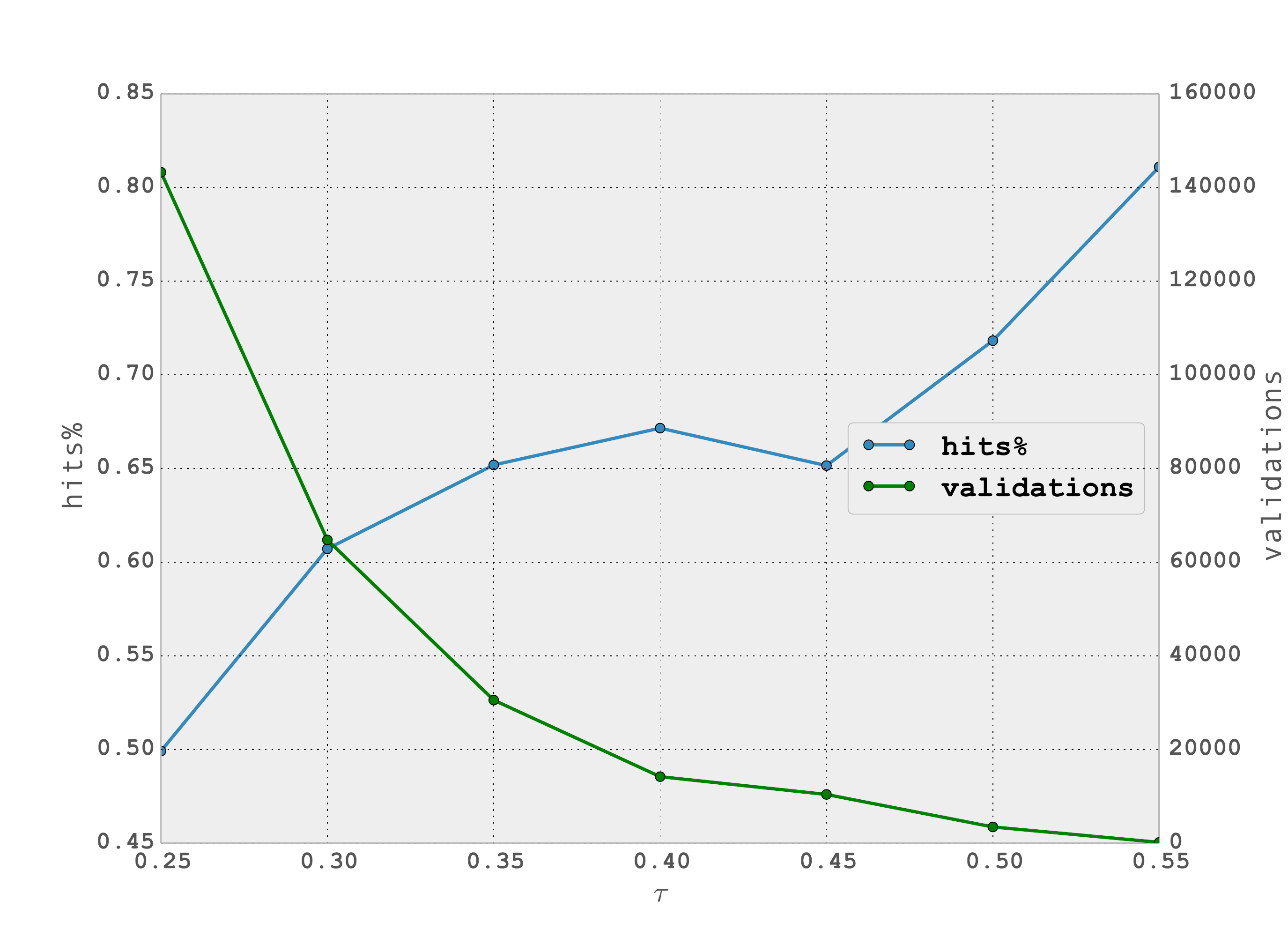}}
	\caption{Performance as function of $\tau$.}
	\label{fig:max_prob_iusacell}
\end{figure}

The focus of our exploratory analysis to find an optimal subset of 
nodes (where our algorithm works best) has been centered on looking 
at the network topology. Here we present a conceptually different 
strategy which exploits the information in the probability vector 
for the age group on each node. Namely, we examine the performance 
of the reaction-diffusion algorithm as we restrict the performance 
to a subset of nodes whose selected category satisfies a minimal 
threshold value $\tau$ in its probability vector. 
In Fig.~\ref{fig:max_prob_iusacell}
we observe a monotonic increase in the performance as the threshold is increased 
and simultaneously a monotonic decrease of the validation set. We note
that for $\tau=0.5$ the performance increased to $72\%$ 
with 3,492 out of the 143,240 (2.4\%) of the validation nodes remaining. 
The performance of the algorithm increases to $\sim 81\%$ for $\tau=0.55$ 
but the number of validation nodes remaining has decreased to 201 ($0.1\%$) 
making this estimate very unreliable.

%% file: conclusion.tex
\section{Conclusion and Future Work} \label{sec:conclusion}

Large scale social networks such as those emerging from mobile phone communications are
rapidly increasing in size and becoming more ubiquitous around the globe. Understanding
how these community structures forming complex network topologies can inform mobile
phone operators, as well as other organizations, on unknown attributes of their users is of
vital importance if they are to better understand, not only their clients' interests and behavior, but
also their social environment including the
users that are not clients of the mobile operator. A better description of this
ecosystem can give the mobile operators a business edge in areas such as churn prediction,
targeted marketing, and better client service among other benefits.

In this work 
we focused our attention on the bare bones
topology of the mobile network, in other words, we aimed at uncovering the potential of the topology
of the network itself to inform us on user attributes, in particular on
user's age groups.
We proposed an algorithm
where the only prior knowledge for inference is
the age group of a subset of the network nodes (the seed nodes), and the
network topology itself. 
Our algorithm
achieved a performance of $46.6\%$ for the entire network and of $53\%$ for the 
age group $35-50$ where pure random selections would have achieved a performance of $25\%$,
or at most $\sim 36\%$ by labeling each node with the most frequent age group.
To understand why the communication topology is allowing for this increase in performance
we first studied the age correlations of users communicating between each other where
we included the labeled set.
As shown in Fig.~\ref{fig:communications}, we can observe a strong age homophily
between callers which is being harnessed by our algorithm.

We next searched for basic
topological features to guide us in finding a subset of nodes where our
algorithm performs best.
Optimizing over all three metrics described in section~\ref{sec:performance-metrics},
we found a subset of $20,050$ nodes where the performance of our algorithm increased 
to $62\%$.

Given the particular nature of the mobile graph, 
where client nodes and non clients nodes stand on a different footing, 
we tested the reaction-diffusion algorithm on the client
only subgraph to examine differences in performance. Even  though
performance increased somewhat, relations of the topological metrics with performance behaved quite similarly and we were
able to again find a sweet spot subset of nodes in $\calN_C$, characterized by the topological metrics, for which the
performance of our algorithm increased to $\sim 67\%$.

Next we showed that the performance of the reaction-diffusion algorithm is robust to changes 
in the parameter $\lambda$ which governs the relative importance of the diffusion and reactive term, as long as both these terms are present. We also saw the performance reaches a stationary value after 5 iterations.
We found that restricting the prediction over nodes with stronger inference given
by their probability vector, we could achieve an increase of up to $72\%$ in performance 
for a significant subset of nodes.

In conclusion, in this work we have presented a novel algorithm that can harness the bare bones topology of mobile phone
networks to infer with significant accuracy the age group of the network's users.
We have shown the importance
of understanding nodes topological properties, in particular their relation to the seed nodes,  
in order to fine grain our expectation of correctly classifying the nodes.
Though we have carried out this analysis for a specific network using a particular algorithm, 
we believe this approach can be useful to study
graph based prediction algorithms in general.

As future work, one direction that we are investigating
is to combine the graph based inference approach presented here
with classical machine learning techniques based on node features \cite{sarraute2014}.  
We are also interested in applying our methodology 
to predict variables related to the users' spending behavior.
In \cite{singh2013predicting} the authors show correlations between social features
and spending behavior for a small population (52 individuals).
We are currently tackling the problem of predicting spending behavior characteristics
on the scale of millions of individuals.

Another research direction is to study the correlations between
the user's mobility and their demographic attributes.
The geolocation information contained in the CDRs
allows the analysis of users' mobility patterns,
e.g. their mobility motifs \cite{schneider_unravelling_2013}.
Recent studies have focused on the mobility patterns related to cultural events,
such as sports \cite{ponieman2013human,xavier2013understanding},
where differences between genders and age groups may naturally be found.
We expect that the interplay between mobility patterns and demographic attributes (such as gender and age)
will provide relevant features to feed our prediction algorithms, 
and more generally will contribute to gain further insights on the behavior of different 
segments of the population.